\documentclass[12pt]{article}

\usepackage{amsmath,amssymb}

\usepackage[T2A]{fontenc}
\usepackage[cp1251]{inputenc}

\usepackage[english,russian]{babel}
\usepackage[all]{xy}

\allowdisplaybreaks[3]

\addto\extrasrussian{}

\makeatletter
\renewcommand{\@cite}[2]{%
{\rm[}{{\rm#1}\if@tempswa , #2\fi}{\rm]}}
\renewcommand{\@biblabel}[1]{\hfill #1. \hspace{-0.4em}}
\makeatother

\newcounter{sectiona}
\newcommand{\sectiona}[1]{\refstepcounter{sectiona}
\begin{center}
\arabic{sectiona}.\ #1
\end{center}}

\newcounter{theorema}
\renewcommand{\thetheorema}{
\arabic{theorema}}
\newcommand{\theorema}[1]{\refstepcounter{theorema}
\par \textbf{Theorem\, \thetheorema.}\, {\it #1}}

\newcounter{propositiona}[sectiona]

\newcounter{remarka}
\renewcommand{\theremarka}{
\arabic{remarka}}
\newcommand{\remarka}[1]{\refstepcounter{remarka}
\textbf{Remark \theremarka.}\, #1}

\newcounter{examplea}
\renewcommand{\theexamplea}{
\arabic{examplea}}
\newcommand{\examplea}[1]{\refstepcounter{examplea}
\textbf{Example \theexamplea.}\, #1}

\newcounter{definitiona}

\begin{document}

\baselineskip 5.8mm

\begin{center}
\bf \Large On the correspondence between variational principles\\
in Eulerian and Lagrangian descriptions
\end{center}

\centerline{\bf\large Alexander V.~Aksenov$^{*,**,1}$ and Konstantin~P.~Druzhkov$^{*,***,2}$}

{\it
\begin{center}
$^*$Lomonosov Moscow State University, Mechanics and Mathematics Department,\\
Leninskie Gory 1, Moscow, 119234 Russia\\
$^{**}$National Research Nuclear University MEPhI,\\
31 Kashirskoe Shosse, Moscow, 115409 Russia\\
$^{***}$Moscow Institute of Physics and Technology (National Research University),\\
Dolgoprudny, Moscow Region, 141700 Russia\\
E-mail: $^1$Aksenov.AV@gmail.com, $^2$Konstantin.Druzhkov@gmail.com
\end{center}
}

\textbf{Abstract.} A relation between variational principles for equations of continuum
mechanics in Eulerian and Lagrangian descriptions is considered. It
is shown that for a system of differential equations in Eulerian
variables corresponding Lagrangian description is related to
introducing nonlocal variables. The connection between these descriptions is obtained in terms of differential
coverings. The relation between variational principles of a system
of equations and its symplectic structures is discussed. It is shown
that if a system of equations in Lagrangian variables can be derived
from a variational principle then there is no corresponding
variational principle in Eulerian variables.

\sectiona{INTRODUCTION}

In continuum mechanics  is well known, that for a given system of
equations in Eulerian variables there exists natural analogue in
Lagrangian variables, which describes the same continuum motion. The
connection between the initial system and its analogue in Lagrangian
variables is given by a differential covering~\cite{VinKr}. From the
Eulerian point of view, Lagrangian description contains nonlocal
variables, which can affect on all main geometrical structures,
including symmetries, conservation laws (see e.g.~\cite{AksDr}),
symplectic and Hamiltonian structures as well.

There are different kinds of variational principles in continuum
mechanics (see e.g.~\cite{Berd}). In our paper,  we shall mean a stationary-action principle by a variational
principle. Action
functional corresponds to some differential form (Lagrangian) $L$
$$
\mathcal{L} = \int\! L
$$
and its stationary points are solutions of the corresponding
Euler-Lagrange equation $\mathrm{E}(L) = 0$,
where $\mathrm{E}$ is the Euler operator.

The paper~\cite{BamMor} is also devoted to the problem of connection
between variational principles in Eulerian and Lagrangian
descriptions. In this paper the authors used nonlocal
variables and hence they deal with some intermediate description
instead of purely Eulerian one. In our paper we obtain the relation
between variational principles in Eulerian description
(without nonlocal variables) and Lagrangian one. The relation is
based on the concept of a symplectic structure for a system of
differential equations.

\vfill\eject

\sectiona{LAGRANGIAN DESCRIPTION AS DIFFERENTIAL COVERING}

Let us consider the mass conservation law in continuum mechanics
\begin{equation}
\rho_t + (u\rho)_x + (v\rho)_y + (w\rho)_z = 0\,.
\label{massc}
\end{equation}
Here $\mathbf{v} = u\partial_x + v\partial_y + w\partial_z$ is a
velocity field, $\rho$ is a mass density. Further we assume that
$\rho > 0$.

Choosing suitable
nonlocal variables, one can introduce a potential for the mass
conserva\-tion law, which satisfy the following relation
\begin{equation}
\begin{aligned}
\rho\, dx\wedge dy\wedge dz &- u\rho\, dt\wedge dy\wedge dz +
v\rho\, dt\wedge dx\wedge dz -{}\\
&- w\rho\, dt\wedge dx\wedge dy = d(\xi^1 d\xi^2\wedge d\xi^3) =
d\xi^1\wedge d\xi^2\wedge d\xi^3\,.
\end{aligned}
\label{poten}
\end{equation}
The relation~\eqref{poten} is equivalent to the following system of
equations
\begin{equation}
\begin{aligned}
&\rho = \det\Big(\dfrac{\partial \xi}{\partial x}\Big)\,,
\qquad \xi^i_t + u\xi^i_x + v\xi^i_y + w\xi^i_z =
0,\qquad i = 1, 2, 3\,.\\
\end{aligned}
\label{cond}
\end{equation}
Here by $\partial\xi/\partial x$ we denote the corresponding Jacobi
matrix. Note that such functions\\ $\xi^i(t,x,y,z)$ are
Lagrangian variables.

\remarka{Transformations that preserve the volume form in $(\xi^1,
\xi^2, \xi^3)$-space form symmetry group for equations in Lagrangian
variables.}

Since the only consistency condition for the system~\eqref{cond} is the
mass conservation law~\eqref{massc}, then for any system of
equations in Eulerian variables the potential~\eqref{poten}
determines its differential covering. If a system of equations at
hand is of the form (in Eulerian variables)
\begin{equation}
\begin{aligned}
&F^1 = 0\,,\qquad \ldots\,,\qquad F^m = 0\,,
\end{aligned}
\label{Euls}
\end{equation}
then the mass conservation law allows us to
derive the covering system in the following form
\begin{equation}
\begin{aligned}
&F^1 = 0\,,\qquad \ldots\,,\qquad F^m = 0\,,\\
&\rho = \det\Big(\dfrac{\partial \xi}{\partial x}\Big)\,,
\qquad \xi^i_t + u\xi^i_x + v\xi^i_y + w\xi^i_z = 0\,,\qquad i=1,2,3\,.
\end{aligned}
\label{cover}
\end{equation}
One can choose $x, y, z$ as new dependent variables
for~\eqref{cover} and obtain usual Lagrangian representation of the
system~\eqref{Euls}. Therefore, for a system of equations in
Eulerian variables the corresponding Lagrangian description can be
considered as differential covering. This fact allows to lift
some geometrical structures from the Eulerian description to the
Lagrangian one.

\remarka{Other conservation laws of systems of equations also allow to
introduce nonlocal variables and to obtain differential coverings.}

\sectiona{INFINITE PROLONGATION OF A SYSTEM OF DIFFERENTIAL EQUATIONS}

Lets consider a system of differential equations
\begin{equation}
\begin{aligned}
&F^1 = 0\,,\qquad \ldots\,,\qquad F^m = 0\,,
\end{aligned}
\label{Sys}
\end{equation}
where $F^i$ are functions of independent variables $x^1$, $...$,
$x^n$, dependent variables $u^1$, $...$, $u^m$ and derivatives
up to some finite order. Denote multi-index of the form $\alpha_i
x^i$ by $\alpha$. Here and further we assume summation over repeated
indices. Put
\begin{equation*}
D_{\alpha} = D_{\!x^1}^{\ \alpha_1}\circ\ldots\circ D_{\!x^n}^{\
\alpha_n}\,,\qquad u^i_{\alpha} = D_{\alpha}(u^i)\,,
\end{equation*}
where $D_{x^i}$ are the operators of total derivatives.
Denote the infinite prolongation of the system of equations $F=0$ by
$\mathcal{E}$
\begin{equation*}
\mathcal{E}\colon\quad F^i = 0\,,\quad  D_{x^1}(F^i) = 0\,,\quad
D_{x^2}(F^i) = 0\,,\quad \ldots
\end{equation*}

\noindent
The universal linearization operator~\cite{VinKr} for $F$ denote by $l_{F}$.
It acts on a
vector-function $\varphi = (\varphi^1, ..., \varphi^m)^T$ by the formula
\begin{align*}
(l_F(\varphi))^i = {l_F}^{i}_{j}(\varphi^j) =
\dfrac{\partial F^i}{\partial u^j_{\alpha}}\, D_{\alpha}(\varphi^j)\,.
\end{align*}
Denote by $l_{\mathcal{E}}$ the restriction of the operator $l_F$ to
the system $\mathcal{E}$.

Further we deal with $l$-normal~\cite{VinKr} systems only.
We shall say that a system of equations of the form
\begin{equation}
\begin{aligned}
&u^1_{\,b_1x^n} = \varPhi^1\,,\qquad \ldots\,,\qquad
u^m_{\, b_mx^n} = \varPhi^m\\
\end{aligned}
\label{Kovf}
\end{equation}
has the extended Kovalevskaya form if all $b_i$ are
positive integers and the right-hand side $(\varPhi^1,\ldots,
\varPhi^m)^T$ is independent of the variables
$u^i_{\,b_ix^n}$ and their derivatives.
Systems of equations, which can be written in an extended Kovalevskaya form, are
$l$-normal. Most systems of equations in continuum mechanics can be
written in an extended Kovalevskaya form.

\examplea{In dimensionless variables the Navier-Stokes system
of equations for incomp\-ressible fluid
\begin{equation*}
\begin{aligned}
&u_t + uu_x + vu_y + wu_z = -p_x + u_{xx} + u_{yy} + u_{zz}\,,\\
&v_t + uv_x + vv_y + wv_z = -p_y + v_{xx} + v_{yy} + v_{zz}\,,\\
&w_t + uw_x + vw_y + ww_z = -p_z + w_{xx} + w_{yy} + w_{zz}\,,\\
&u_x + v_y + w_z = 0
\end{aligned}
\end{equation*}
can be written in the extended Kovalevskaya form for $x^n=z$,
$b=(2,2,1,1)$. One can eliminate $w_{z}$ from the fourth
equation; eliminate $u_{zz}$ from the first equation; eliminate
$v_{zz}$ from the second equation; eliminate $p_{z}$ from the third
equation. The Euler system of equations also can be written in an
extended Kovalevskaya form.}

\sectiona{SYMPLECTIC STRUCTURES AND VARIATIONAL PRINCIPLES}

For $l$-normal systems of differential equations, similar to the
classical differential geometry, there are two equivalent
representations of symplectic structures: as equivalence classes of
differential forms,
and as equivalence classes of operators in total derivatives  (see, e.g.~\cite{VinKr}).\\
1. A symplectic structure of a system of equations $\mathcal{E}$ is a
closed variational $2$-form on $\mathcal{E}$, i.e. an element of the
kernel of the variational differential
$$
\delta\colon E^{2,\, n-1}_1(\mathcal{E})\to E^{3,\,
n-1}_1(\mathcal{E})\,.
$$
Here $E^{p,\, n-1}_1(\mathcal{E})$ are groups of variational $p$-forms
from the $\mathcal{C}$-spectral sequence.\\
2. Variational $2$-forms of a system $\mathcal{E}$ can
be described as operators in total derivatives $\Delta$,
which satisfy the relation
\begin{equation*}
\Delta^*\circ l_{\mathcal{E}} = l_{\mathcal{E}}^{\, *} \circ \Delta\,,
\end{equation*}
modulo operators of the form $\nabla \circ l_{\mathcal{E}}$, where
$\nabla = \nabla^*$. Here the operator $\Delta^*$ is formally adjoint
to an operator $\Delta$.

Each symplectic structure of an $l$-normal system of differential
equations determines a map from its symmetries to variational
$1$-forms (i.e. it determines a Noether theorem). This map can be
equivalently described in both ways: in terms of variational forms
and in terms of operators. Such operators map symmetries of
$\mathcal{E}$ to $\mathrm{Ker}\, l^{\,*}_{\mathcal{E}}$, which is
isomorphic to the group of variational 1-forms $E^{1,\,
n-1}_1(\mathcal{E})$ in $l$-normal case.

\remarka{The definition of a symplectic structure as a closed
variational $2$-form allows us to lift symplectic structures in
coverings.}

We shall say that a system of equations $\mathcal{E}$ is variational
if there exists a Lagrangian $L$, such that the following relation
holds
\begin{equation*}
F = \mathrm{E}(L)\,,
\end{equation*}
where $\mathrm{E}$ is the Euler operator. Also we shall say that the
system of equations $\mathcal{E}$ admits variational principle if
for some operator in total derivatives $A$ exists a Lagrangian $L$,
such that the following relation holds
\begin{equation*}
A(F) = \mathrm{E}(L)\,.
\end{equation*}
In such situation the operator $\Delta = A^*|_{\mathcal{E}}$
determines the symplectic structure for the correspon\-ding system
$\mathcal{E}$. Let us say that a symplectic structure of
$\mathcal{E}$, which can be obtained in this way, is related to a
variational principle.

\sectiona{MAIN RESULTS}

Informally speaking, symplectic structures of a system of
differential equations can be considered as its "Noether theorems".
Therefore it is quite natural to expect that at least for an
$l$-normal system of differential equations its symplectic
structures can be used for deriving variational principles.

\theorema{
If an $l$-normal system of equations $\mathcal{E}$ has
trivial de Rham cohomology group $H^{n+1}(\mathcal{E})$, then each
symplectic structure of $\mathcal{E}$ is related to a variational
principle.}

\textbf{Proof.} Each symplectic structure can be considered as
equivalence class of differential forms. As it follows
from~\cite{Khavk} a symplectic structure $\omega\in E^{2,\,
n-1}_1(\mathcal{E})$ is related to a variational principle if and only if it is
generated by an exact differential form (as equivalence class of
differential forms). Each closed variational 2-form $\omega$ either
belongs to $\delta E^{1,\, n-1}_1(\mathcal{E})$ or generates
nontrivial element $[\omega]$ of cohomology group $E^{2,\,
n-1}_2(\mathcal{E})$. If $\omega\in \delta E^{1,\,
n-1}_1(\mathcal{E})$, then it is generated by an exact differential
form. Assume now that $[\omega]\neq 0$. According to the two-line
theorem (see~\cite{VinKr}) group $E^{\,4,\, n-2}_2(\mathcal{E})$ is
trivial, hence $[\omega]$ is a cocycle. The group $E^{2,\,
n-1}_3(\mathcal{E})$ is also trivial, then $[\omega] \in
\mathrm{Im}\, d^{\,0,\, n}_2$, where
$$
d^{\,0,\, n}_2\colon E^{0,\, n}_2(\mathcal{E})\to E^{2,\, n-1}_2(\mathcal{E})\,.
$$
Then $[\omega] = \omega + \delta E^{1,\, n-1}_1(\mathcal{E})$ is
generated by an exact differential form. Hence symplectic structure
$\omega$ is still generated by an exact differential form and is
related to a variational principle.

A system of equations in an extended Kovalevskaya form
admits a canonical way to derive a variational principle from a
symplectic structure~\cite{Dr}. Therefore, if a system of equations
is written in an extended Kovalevskaya form in both Eulerian and
Lagrangian descriptions, then the relation between variational
principles in these descriptions can be obtained
in terms of symplectic structures. Each variational principle in
Eulerian description generates unique symplectic structure, which
can be lifted to the Lagrangian description. The last step is to
derive the corresponding variational principle from this lift.

Let $F = 0$ be a system of differential equations in Eulerian
variables, $\mathcal{E}$ be its infinite prolongation,
$\widetilde{\mathcal{E}}$ be the corresponding system in Lagrangian
variables. Denote the covering from $\widetilde{\mathcal{E}}$ to
$\mathcal{E}$ by $\tau$.

\theorema{
If the system of equations $\widetilde{\mathcal{E}}$ is
$l$-normal variational system, then the corresponding symplectic
structure of $\widetilde{\mathcal{E}}$ is not a lift of a symplectic
structure of $\mathcal{E}$.
}

\textbf{Proof.} Consider algebra ${\tau_*}$-$\mathrm{sym}\,
\widetilde{\mathcal{E}}$ of $\tau$-projectable symmetries of
$\widetilde{\mathcal{E}}$. Then for any variatio\-nal 2-form
$\omega\in E^{2,\, n-1}_1(\mathcal{E})$ the following diagram is
commutative
\begin{align*}
\xymatrix
{
{\tau_*}\text{-}\mathrm{sym}\, \widetilde{\mathcal{E}}\ar[d]_{\tau_{\ast}}\
\ar[rr]^{\tau^{\ast}(\omega)\quad} &&\ E^{1,\, n-1}_1(\widetilde{\mathcal{E}})&\\
\mathrm{sym}\, \mathcal{E}\ \ar[rr]^{\omega\quad} &&\ E^{1,\, n-1}_1(\mathcal{E})\ar[u]_{\tau^{\ast}}&\\
}
\end{align*}
Thus, for a symmetry $\varphi\in \tau_*$-$\mathrm{sym}\,
\widetilde{\mathcal{E}}$, which acts in a fibre of $\tau$, holds the
relation $\tau^*(\omega)(\varphi) = 0$, i.e. the lift of a
symplectic structure of $\mathcal{E}$ is degenerate and the
equivalence class of operators, which corresponds to
$\tau^*(\omega)$, can not contain the identical operator.

This theorem shows that if an $l$-normal system of equations in
Lagrangian variables is variational, then the corresponding
variational principle has no analogues in Eulerian variables.
Similar result holds true for any covering from an $l$-normal system
of differential equations, such that the fiber symmetry algebra is
nontrivial. In particular, it holds true for coverings from
$l$-normal systems of differential equations, which based on
introduction of potentials for conservation laws.

\sectiona{EXAMPLES}

Lets consider two examples of variational $l$-normal systems in
Lagrangian variables.

\examplea{Equations of motion a polytropic gas ($p = C\rho^{\gamma}$)
in Lagrangian variables are Euler-Lagrange equations for Lagrangian
\begin{align*}
L = \Big(\dfrac{x_t^2 + y_t^2 + z_t^2}{2}-V-U\Big)dt
\wedge d\xi^1\wedge d\xi^2\wedge d\xi^3\,,
\end{align*}
where $V$ is a potential energy, $U$ is an internal energy. The
corresponding system of equations can be written in the extended
Kovalevskaya form
\begin{align*}
x_{tt} = -\dfrac{\delta (V + U)}{\delta x}\,,\qquad
y_{tt} = -\dfrac{\delta (V + U)}{\delta y}\,,\qquad
z_{tt} = -\dfrac{\delta (V + U)}{\delta z}\,.
\end{align*}
Thus, this system of equations is $l$-normal and there is no
corresponding variational principle for equations of motion a
polytropic gas in (purely) Eulerian variables.}

\examplea{The Green-Naghdi equation in Lagrangian variables can be
derived from variational principle for Lagrangian~\cite{DoKaMe}
\begin{align*}
\begin{aligned}
L = \Big(&\dfrac{x_t^2}{2}\Big(1 + \varepsilon\big(H''^{\,2} +
\dfrac{x_{mm}}{x_m^3}H'' - \dfrac{1}{2x_m}H'''\big)\Big) +{}\\
&+\dfrac{1}{6x_m^4}\big(\varepsilon x_{tm}^2 -
3gx_m^2(2Hx_{mm} + x_m)\big)\Big)dt\wedge dm\,,
\end{aligned}
\end{align*}
where function $H'(x)$ describes the bottom topography, $g$ is the
gravitational acceleration, $\varepsilon$ is a small parameter.
After $\pi/4$-rotation in $(t, m)$-space the corresponding equation
can be written in the extended Kovalevskaya form. Thus, this
equation is also $l$-normal and there is no corresponding
variational principle for the Green-Naghdi equations in Eulerian
variables.}

\vspace{2.0ex}

\centerline{CONCLUSION}

\vspace{1.5ex}

The construction of a symplectic structure for a system of
differential equations allowed to connect variational principles in
Eulerian and Lagrangian variables. The connection is based on the
fact that one can consider Lagrangian variables as nonlocal
variables in a differential covering. Besides, the obtained result
allowed to show that a nondegenerate variational principle in
Lagrangian variables is not related to any variational principle in
Eulerian variables. However, each variational principle in Eulerian
variables is related to some variational principle in Lagrangian
variables. Thus, from this point of view the Lagrangian description
is preferred over the Eulerian one.

\end{document}